\documentclass[12pt, a4paper]{article}
\usepackage[english]{babel}
\usepackage{graphicx}
\usepackage{fullpage}
\usepackage{amsmath}
\begin{document}
\title{\sc A Flexible Bayesian Model for Estimating Subnational Mortality\footnote{This study was made possible by awards from the U.S. National Institute on Aging (R01-AG011552 and R01-AG040245) and from contributions to the Human Mortality Database project from the Canadian Institute of Actuaries. However, any opinions, findings, and conclusions or recommendations expressed in this material are those of the authors alone and do not necessarily represent the official views of the National Institute on Aging and other funders.}}
\author{Monica Alexander\footnote{\texttt{monicaalexander@berkeley.edu}} \\ \emph{University of California, Berkeley} \and Emilio Zagheni\footnote{\texttt{emilioz@uw.edu}} \\ \emph{University of Washington, Seattle} \and Magali Barbieri\footnote{\texttt{magali@demog.berkeley.edu}} \\ \emph{University of California, Berkeley} and \\\emph{Institut national d'\'{e}tudes d'\'{e}mographiques}}

\maketitle

\abstract{Reliable mortality estimates at the subnational level are essential in the study of health inequalities within a country. One of the difficulties in producing such estimates is the presence of small populations, where the stochastic variation in death counts is relatively high, and so the underlying mortality levels are unclear. We present a Bayesian hierarchical model to estimate mortality at the subnational level. The model builds on characteristic age patterns in mortality curves, which are constructed using principal components from a set of reference mortality curves. Information on mortality rates are pooled across geographic space and smoothed over time. Testing of the model shows reasonable estimates and uncertainty levels when the model is applied to both simulated data which mimic US counties, and real data for French \emph{d\'{e}partements}. The estimates produced by the model have direct applications to the study of subregional health patterns and disparities.}

\setlength\parindent{0pt} 
\setlength{\parskip}{\baselineskip}%

\newpage
\section{Introduction}

In order to effectively study health disparities within a country, it is important to obtain reliable subnational mortality estimates to quantify geographic differences accurately. There is a large demand for estimates of small-area mortality as indicators of overall health and well-being, as well as for natural experiments that exploit policy changes at local levels. Reliable mortality estimates for regional populations could help to better understand how place-of-residence and communities can affect health status, through both compositional and contextual mechanisms (Macintyre et al.\ 2002). 

One of the difficulties in producing mortality estimates for subnational areas is the presence of small populations where the stochastic variation in death counts is relatively high. For example, 10 per cent, or around 300, of US counties have populations of less than 5,100, and 1 per cent of counties have less than 1,000 people (USCB 2015). The resulting mortality rates in small areas are often highly erratic and may have zero death counts, meaning the underlying true mortality schedules are unclear. 

The aim of this paper is to formulate a model for estimating mortality rates at the subnational level, across geographic areas with a wide variety of population sizes and death counts. Resulting estimates would be useful for guiding future policy efforts to improve the health of populations and to investigate the historical impact of public health interventions and changes in the structure of local health programs. In this article, we focus on developing the methodology to produce age- and sex-specific mortality rates and the approach is tested on simulated data that mimic US counties, and on real data for French \emph{d\'{e}partements}.  However, the model is flexible enough to be used in a wide range of situations. 

There has been a growing literature in the field of small-area mortality estimation. However, the demand for accurate, reliable and consistent estimates has not yet been met.  The traditional life table approach assumes that deaths $y_{ax}$ in a population in area $a$ at age $x$ are poisson distributed $y_{a,x} \sim \text{Poisson}(P_{a,x}\cdot m_{a,x})$ where $P_{a,x}$ is the population at risk and $m_{a,x}$ is the mortality rate. The maximum likelihood estimate of the mortality rate for area $a$ at age $x$ is 
\begin{equation*}
\hat{m}_{ax} = \frac{y_{a,x}}{P_{a,x}}. 
\end{equation*}
This approach essentially involves estimating as many fixed-effect parameters $m_{ax}$ as there are data points. In addition, this estimation process makes no reference to or use of the information about mortality rates at other ages, in other areas or at other time points. Confidence intervals can be derived based on the distribution of deaths, but for small populations, stochastic variation is high and so confidence intervals and standard errors are large. Estimating mortality rates in small populations therefore requires different types of approaches.

To avoid issues that arise in small-area mortality estimation, a common approach is to aggregate mortality data across multiple years or across space to form larger geographic areas. For example, recent work by Chetty et al.\ (2016) and Currie and Schwandt (2016) look at mortality inequalities in the US using deaths and income measures at the county level. However, data are either aggregated across space and time (Currie and Schwandt 2016) or results are not published for smaller populations (in the case of Chetty et al.\ (2016), where the minimum population size is 20,000). However, given the information lost in aggregating data from smaller areas, there is value in employing other techniques to infer mortality levels and trends. 

One option that has been employed is to treat each small population as a stand-alone population and model accordingly using traditional model life table approaches. For example, Bravo and Malta (2010) outline an approach for estimating life tables in small populations, applied to regional areas of Portugal. They estimate Gompertz-Makeham functions via generalized linear models, with an adjustment at older ages. Another approach by Jarner and Kryger (2011) involves estimating old-age mortality in small populations by first estimating parameters of a frailty model using a larger reference population. However, approaches that treat small populations separately do not account for the likely relationships between the regional population estimates or patterns over time. Other approaches in the US have used county-level covariates such as socio-economic status and education to predict county-level life expectancy (Ezzati et al.\ 2008; Srebotnjak et al.\ 2010; Kulkarni et al.\ 2011; Kindig and Cheng 2013). However, there are issues with using the resulting estimates to infer relationships between health, poverty rates and education without concerns of endogeneity.
   
In this paper, we propose a new model that relies on a Bayesian hierarchical framework which allows information on mortality to be shared across time and through space. This helps to inform the mortality patterns in smaller geographic areas, where uncertainty around the data is high. The modeling process produces uncertainty intervals around the mortality estimates, which can then be translated into uncertainty around other life-table quantities (for example, life expectancy). As well as producing uncertainty intervals around the final estimates, the modeling process also involves the estimation of other meaningful variance parameters that may relate to variation in mortality within, or across, states.

The remainder of the paper is structured as follows. In the next section, the methodology for estimating age-specific mortality rates is described. The model is then applied to two different data situations -- simulated and real -- and results are discussed. Performance of the model is evaluated through coverage and mean-squared-error measures. The paper finishes with a discussion, including plans for future work. 

\section{Method}
We propose a model that has an underlying functional form that captures regularities in age patterns in mortality. We then build on this functional form within a Bayesian hierarchical framework, penalizing departures from the characteristic shapes across age, as well as sharing information across geographic areas and ensuring a relatively smooth trend in mortality over time. 

Bayesian hierarchical models have previously been used in a wide range of demographic applications. For example, a model proposed by Raftery et al.\  (2012) is used by the United Nations Population Division to produce probabilistic population projections. Alkema and New (2014) developed a Bayesian hierarchical model for estimating the under-five mortality rate for all countries worldwide. There are many other examples in the fields of mortality, fertility and migration (e.g. Condgon 2009; King and Soneji 2011; Sharrow et al.\ 2013; Alkema et al.\ 2012; Bijak 2008). Our approach has similarities to applications by authors in cause of death mortality estimation (Girosi and King 2008) and cohort fertility projection (Schmertmann et al.\ 2014), but with a focus on addressing small-area estimation issues, rather than forecasting.

\subsection{Model set-up}

Let $y_{x,a,t}$ be the deaths at age $x$ in area $a$ at time $t$. We assume that deaths are Poisson distributed 
\begin{equation} \label{deaths}
y_{x,a,t} \sim \text{Poisson}(P_{x,a,t} \cdot m_{x,a,t}), 
\end{equation}
where $m_{x,a,t}$ is the mortality rate at age $x$, area $a$ and time $t$ and $P_{x,a,t}$ is the population at risk at age $x$, area $a$ and time $t$. We estimate mortality rates at ages 0, 1, 5, and then in 5-year intervals up to 85+. 

The $m_{x,a,t}$ are modeled on the log-scale as:
\begin{equation*}
\log(m_{x,a,t}) = \beta_{1,a,t} \cdot Y_{1x} + \beta_{2,a,t} \cdot Y_{2x} + \beta_{3,a,t} \cdot Y_{3x} + u_{x,a,t},
\end{equation*}
where $Y_{px}$ is the $p$th principal component of some set of standard mortality curves, and $u_{x,a,t}$ is a random effect. The use of principal components has similarities with the Lee-Carter approach (Lee and Carter 1992). Principal components create an underlying structure of the model in which regularities in age patterns of human mortality can be expressed. Many different kinds of shapes of mortality curves can be expressed as a combination of the components. Incorporating more than one principal component allows for greater flexibility in the underlying shape of the mortality age schedule. 

Principal components are obtained via a singular value decomposition on a set of standard mortality curves. For example, for the application to simulated US counties below, we used US state mortality rates from 1980--2010. Let $\bf{X}$ be a $N \times G$ matrix of log-mortality rates, where $N$ is the number of observations and $G$ is the number of age-groups. In the US states case, we had $N = 50 \times 31 = 1550$ observations of $G =19$ age-groups. The singular value decomposition of $\bf X$ is
\begin{equation*}
\bf X = \bf{UDV'},
\end{equation*}
where $\bf V$ is a $G \times G$ matrix. The first three columns of $\bf V$ (the first three right-singular values of $\bf X$) are $Y_{1x}, Y_{2x}$ and $Y_{3x}$.\footnote{Throughout the paper, we refer to the $Y_{px}$'s as principal components for simplicity, even though technically $Y_{px}$ is really the $p-$th vector of principal component loadings.} The first three principal components for US state mortality curves from 1980--2010 are shown in Figure \ref{pcs} below. They were based on mortality curves on the log scale. Broadly, the first principal component describes the overall mortality curve. The second principal component allows mortality at younger ages to be higher in relation to adult mortality. The third principal component allows adult mortality to be higher in relation to mortality at young and old ages. For example, in some regions of a country, child mortality might be relatively higher than the baseline schedule. In other regions, where prevalence of deaths due to accidents is higher, adult mortality would be higher than the baseline pattern. The components capture overall patterns of mortality well; a wide range of different mortality curves can be expressed as a linear combination of these three components.

\begin{figure}[h!]
\caption{Principal components of (logged) US state mortality schedules, Males, 1980--2010.}
\label{pcs}
\centering
\includegraphics[width=1.12\textwidth]{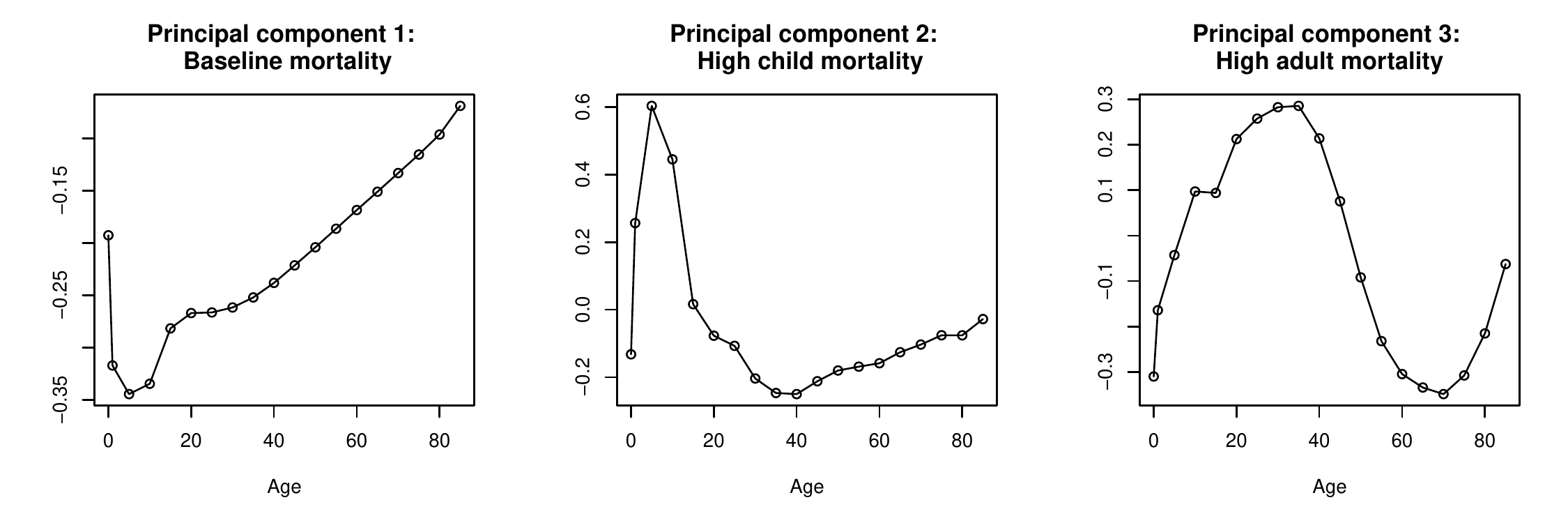}
\end{figure}

The addition of the random effect term $u_{x,a,t}$ in the expression for $\log(m_{x,a,t})$ accounts for potential over-dispersion of deaths, i.e.\ the case where the variance in deaths is greater than the mean, which otherwise would not be expected given the assumption of Poisson-distributed deaths (Congdon 2009). It is assumed that these random effects are centered at zero with an associated variance:
\begin{equation*}
u_{x,a,t} \sim N(0, \sigma^2_x).
\end{equation*}
The variance parameter varies by age group, allowing heterogeneity in some age groups to be greater than in others. In practice it is often the younger age groups, with the lowest levels of mortality, that have the highest variation. 

\subsection{Pooling information across geographic area}
To allow for information on the level and shape of mortality to be shared across geographic space, we assume that the coefficients $\beta_{p,a,t}$  for a particular area are drawn from a common distribution centered around a state (or country) mean:
\begin{eqnarray*}
\beta_{p,a,t} &\sim& N(\mu_{\beta_{p,t}}, \sigma^2_{\beta_{p,t}})
\end{eqnarray*}
where $p$ indicates principal component ($p=1,2,3$). Larger areas in terms of population size (and death counts) have a bigger effect on the overall means. The less data available on deaths in an area, the closer the parameter estimates are to the mean parameter value. In this way, mortality patterns in smaller areas are partially informed by mortality patterns in larger areas. At the same time, mortality patterns in larger areas borrow little information from the pooling process and are largely determined by their own observed death counts. 

\begin{figure}[h!]
\caption{Illustrating the effect of geographic pooling.}
\label{pooled}
\centering
\includegraphics[width=0.9\textwidth]{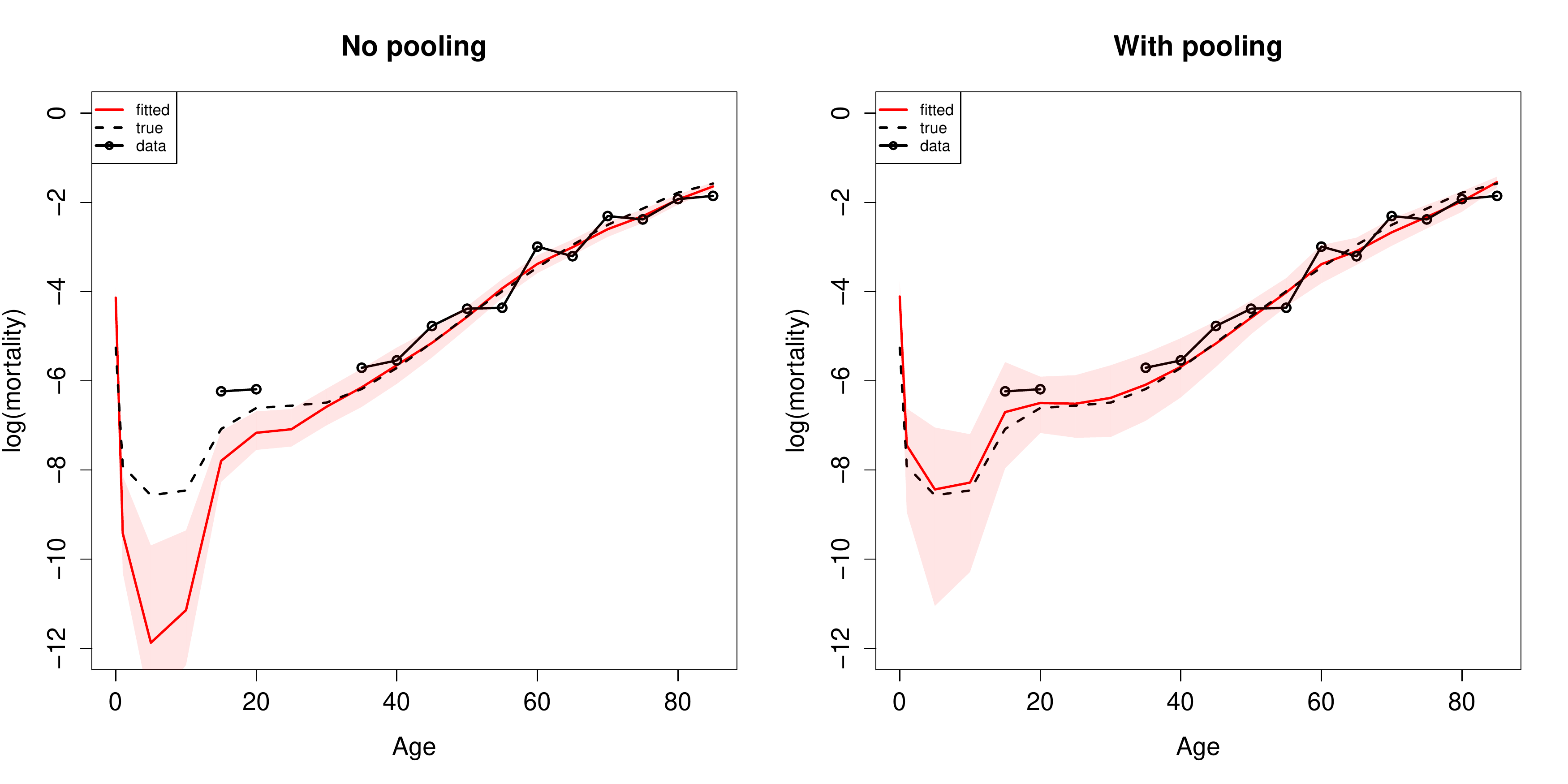}
\end{figure}

The influence of the geographic pooling is illustrated in Figure \ref{pooled}. The charts illustrate observed, true and fitted log-mortality rates for a hypothetical county with a population of 5,000 males. The black dashed line is the true underlying log-mortality rate. The black dots represent the observed log-mortality rates; these were simulated from the true rates using equation \ref{deaths}. Where there are gaps, the observed death count was zero. The red line and associated shaded area is the fit and 95 per cent credible intervals. The graph on the left showed a fit without geographic pooling, while the graph on the right shows a fit with geographic pooling. The effect of pooling is seen most in log-mortality rates at younger ages, where rates are low. As many of the younger age groups have observed zero death counts, the unpooled model estimates log-mortality rates that are much lower than the true rates. The pooled model benefits from information on young-age mortality from other counties, giving a more reasonable estimate. 

In practice, the mean parameter values $\mu_{\beta_{p,t}}$ could be determined from any plausible group of areas which share similar characteristics. We have tested the model using state level mean parameter values, but other options include grouping areas by a smaller location scale, by rural/urban area within state, or by a common age distribution. 

\subsection{Smoothing across time}
We assume the parameters governing the shape of the mortality curve,  $\beta_{p,a,t}$, change gradually and in a relatively regular pattern over time. We impose this smoothing by penalizing the second-order differences across time in the mean parameters:
\begin{eqnarray*}
\mu_{\beta_{p,t}} &\sim& N\left( 2\cdot \mu_{\beta_{p,t-1}} - \mu_{\beta_{p,t-2}}, \sigma^2_{\mu_{\beta_{p,t}}} \right)\\
\end{eqnarray*}
for $p=1,2,3$. This set up is penalizing differences from a linear trend in the mean parameters. Smoothing the mean parameters, rather than the actual parameters $\beta_{p,a,t}$, still allows for mortality trends to depart from a smooth trajectory if suggested by the data. For example, if a particular area suffered from a influenza outbreak thus making mortality higher than in previous years, the $\beta_{p,a,t}$ terms would allow for higher mortality. 

\begin{figure}[h!]
\caption{Illustrating the effect of smoothing over time.}
\label{smoothed}
\centering
\includegraphics[width=0.6\textwidth]{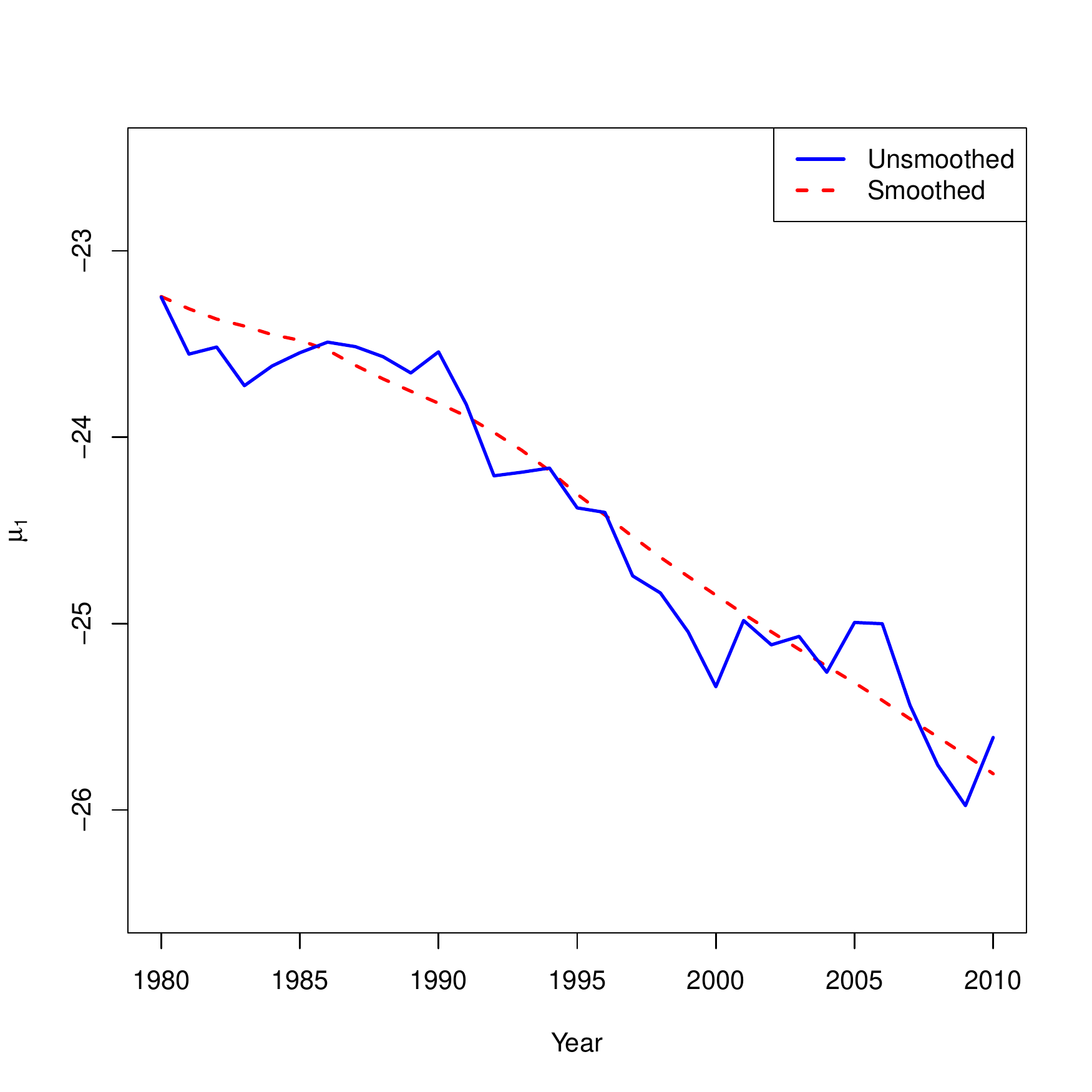}
\end{figure}

The effect of smoothing parameters over time is shown in Figure \ref{smoothed}. The graph shows the estimated median value of the parameter $\mu_{\beta_1}$ over 31 years in a simulated US county model. The blue line shows the estimates without smoothing, while the red dashed line shows the effect of smoothing. 

\subsection{Adding constraints to the model for total areas}
While mortality rates are estimated for subnational populations, it is important that these mortality rates, when aggregated to the state or national level, are consistent with the mortality rates observed at the aggregate level. To ensure this is the case, we add a constraint to the model which specifies that the number of deaths in a state (or country) is Poisson distributed with a rate equal to the sum of all estimated deaths in all areas:
\begin{equation*}
\sum_{a=1}^A y_{x,a,t} \sim \text{Poisson}\left(\sum_{a=1}^A (P_{x,a,t} \cdot m_{x,a,t})\right)
\end{equation*}
where $A$ is the total number of areas.

\subsection{Priors and Implementation}
Non-informative priors were put on variance parameters. Operationally, we used a uniform distribution between 0 and 40 for the standard deviations: 
\begin{eqnarray*}
\sigma_{\beta_{p,t}} &\sim& U(0,40)\\
\sigma_{\mu_{\beta_{p,t}}} &\sim& U(0,40)\\
\sigma_x &\sim& U(0,40).
\end{eqnarray*}

The model was fitted in a Bayesian framework using the statistical software R. Samples were taken from the posterior distributions of the parameters via a Markov Chain Monte Carlo (MCMC) algorithm. This was performed through the use of JAGS software, an R package developed by Plummer (2003). Standard diagnostic checks using trace plots and the Gelman and Rubin diagnostic (Gelman and Rubin 1992) were used to check convergence.

\subsection{Simulation of data for model testing}

In order to test the model, we created a simulated data set of deaths and populations that mimic counties within US states. The `true' mortality rate in a county is based on a specified population and age structure, and the mortality rate in the state. The mortality curve for a county can be altered to change shape via a Brass relational model setup, assuming that:
 \begin{equation} \label{brass}
\log \left ( \frac{l_{x}}{1- l_{x}} \right ) = \alpha + \beta \cdot Y_x
\end{equation}
where $l_x$ is the survivorship at age $x$ and $Y_x$ is the survivorship at age $x$ in the state of interest. To alter the shape of the survivorship curve for a particular county, the values of $\alpha$ and $\beta$ were changed. The values of $\alpha$ and $\beta$ were chosen randomly from the ranges $[-0.75,0.75]$ and $[0.7, 1.3]$, respectively. These ranges of $\alpha$ and $\beta$ values were chosen because they translate to a reasonable range of age-specific mortality curves commonly observed. The survivorship rates were then converted to mortality rates using standard life table relationships. 

Once the `true' mortality rate schedules were obtained, we simulated deaths according to the relationship shown in Equation \ref{deaths}. A range of population sizes were tested, with the minimum county size being 1,000 people of a particular sex. At this small population size, many simulated death counts for particular age groups are equal to zero.

\section{Results}

\subsection{Simulated data}
\begin{figure}[h!]
\caption{True, simulated and estimated log-mortality rates of three counties. }
\label{simulated}
\centering
\includegraphics[width=1.15\textwidth]{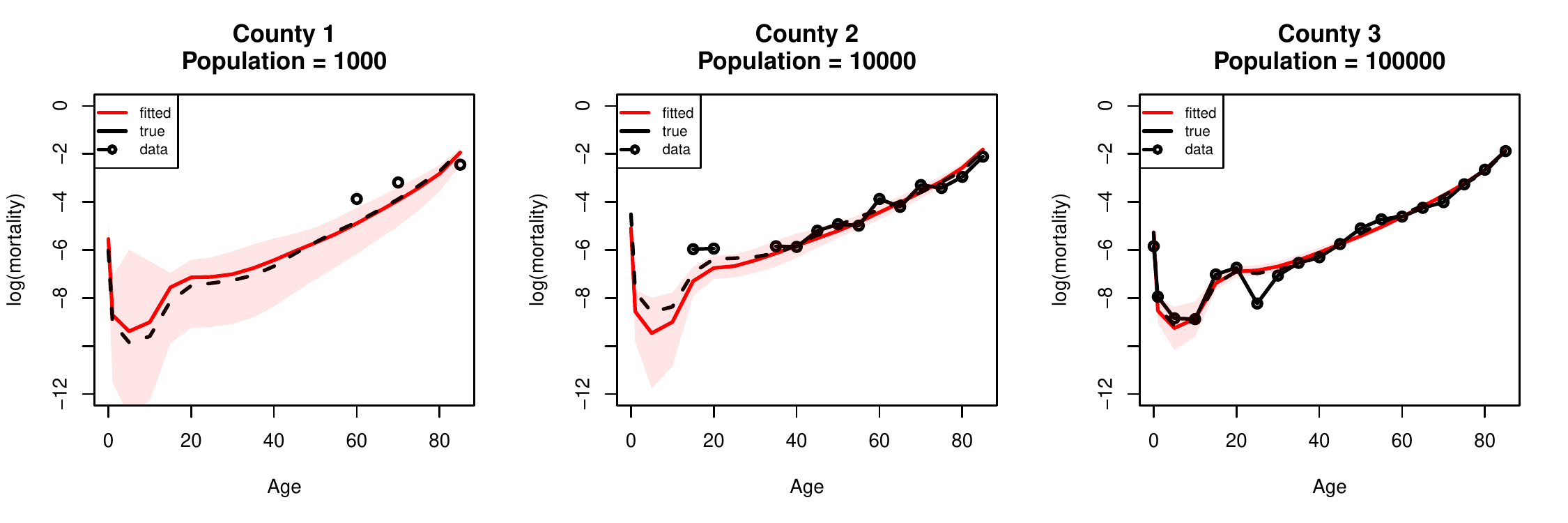}
\end{figure}

Figure \ref{simulated} shows the true, simulated (`observed') data and estimated mortality rates on the log scale in three hypothetical counties within the same state but with different population sizes.  The points show the observed data, which is simulated from the true underlying mortality rate, shown by the black dashed line. For the smallest county, which has 1,000 people, many of the observed death counts are zero, so the data do not show up on the log scale.  The red line shows the estimated log-mortality rates, and the corresponding red shaded area shows the 95 per cent credible intervals.  As the size of the county increases, the mortality pattern in the observed data becomes more regular. As such, the uncertainty around the estimates decreases with increased population size. 

\subsubsection{Evaluation of model performance}
In order to evaluate model performance, we compared the model fit to the fit of a simple Loess smoother and Brass model. The Loess approach does not incorporate any pooling of information or demographic regularities across age. The Brass estimation process uses Equation \ref{brass}. The methods of estimation were compared using the simulated data, where the true value of the mortality rates was known. For each area and time, we estimated the root mean squared error (RMSE), defined as
\begin{equation*}
\text{RMSE} = \sqrt{\sum_{i=1}^G(\hat{m}_{x} - m_{x}^*)},
\end{equation*}
where $\hat{m}_x$ is the estimated mortality rate at age $x$, $m_{x}^*$ is the true mortality rate and $G$ is the number of age groups. 

Table \ref{rmse} below shows the average RMSE for the three fits to a simulated dataset containing 60 counties over 31 years (12 counties per size group). In all cases, the RMSE decreases as county size increases. This is intuitive because, as the county population increases, there are fewer zero death counts and so more information about the shape of the mortality curve. The average RMSE for the model is always lower than Loess or Brass, irrespective of county size. Although the Brass RMSE seem reasonable, it is most likely because the data were generated using a Brass relational model. 

\begin{table}[h!]
\centering
\caption{Average RMSE for Model, Loess and Brass Fits}
\label{rmse}
\begin{tabular}{r|rrl}
            & \multicolumn{3}{c}{RMSE} \\ \hline
County Size & Model   & Loess  & Brass  \\ \hline
1,000       & 0.034   & 0.187  & 0.039  \\
5,000       & 0.027   & 0.078  & 0.037  \\
10,000      & 0.027   & 0.065  & 0.042  \\
20,000      & 0.022   & 0.052  & 0.030  \\
100,000     & 0.013   & 0.049  & 0.027 
\end{tabular}
\end{table}

In addition, we also evaluated the relationship between the nominal and actual coverage for the uncertainty intervals produced by the Bayesian model. Coverage is defined as:
\begin{equation*}
\frac{1}{G} \sum_{i=1}^G 1\left[m_{x}^* \geq l_{x}\right] 1\left[m_{x}^*< r_x\right]
\end{equation*}
where $G$ is the number of age groups, $m_{x}^*$ is the true mortality rate for the $x$th age group and $l_x$ and $r_x$ the lower and upper bounds of the credible intervals for the $x$th age group. Coverage at the 80, 90 and 95 per cent levels was considered. Table \ref{coverage} shows the average coverage for the proposed model, fit to 60 counties over 31 years. In general, actual coverage levels are close to the nominal level, indicating that the model is well-calibrated. The coverage level tends to decrease as county size increases. 

\begin{table}[h!]
\centering
\caption{Nominal versus actual coverage}
\label{coverage}
\begin{tabular}{r|rrr}
            & \multicolumn{3}{c}{Coverage level (\%)} \\ \hline
County Size & 80         & 90         & 95        \\ \hline
1,000       & 0.871      & 0.952      & 0.982     \\
5,000       & 0.799      & 0.896      & 0.940     \\
10,000      & 0.749      & 0.853      & 0.890     \\
20,000      & 0.744      & 0.833      & 0.891     \\
100,000     & 0.763      & 0.865      & 0.901    
\end{tabular}
\end{table}


\subsection{Application to French \emph{D\'{e}partements}}
We also tested the model on real mortality data, applied to death and population counts by sex in French \emph{d\'{e}partements} from 1975--2008 (INSEE 2015).\footnote{We chose to use French \emph{d\'{e}partements} data because, at the time of writing, data for all US counties was not readily available.} The annual life tables were constructed by the \emph{Division des statistiques r\'{e}gionales, locales et urbaines} [Regional, Local and Urban Statistics Division] of the French \emph{Institut national de la statistique et des \'{e}tudes \'{e}conomiques} or INSEE [National Institute for Statistics and Economic Studies]. The life tables were built from the vital statistics and census data also collected and processed by INSEE. There are 96 French \emph{d\'{e}partements} ranging in population size from around 35,000 to 1.5 million (for one sex). 

\begin{figure}[h!]
\centering
\caption{Observed and estimated log-mortality rates, Loz\`{e}re, Males, 1975 and 2008.}
\label{MEL}
\begin{tabular}{cc}
\includegraphics[width=0.9\textwidth]{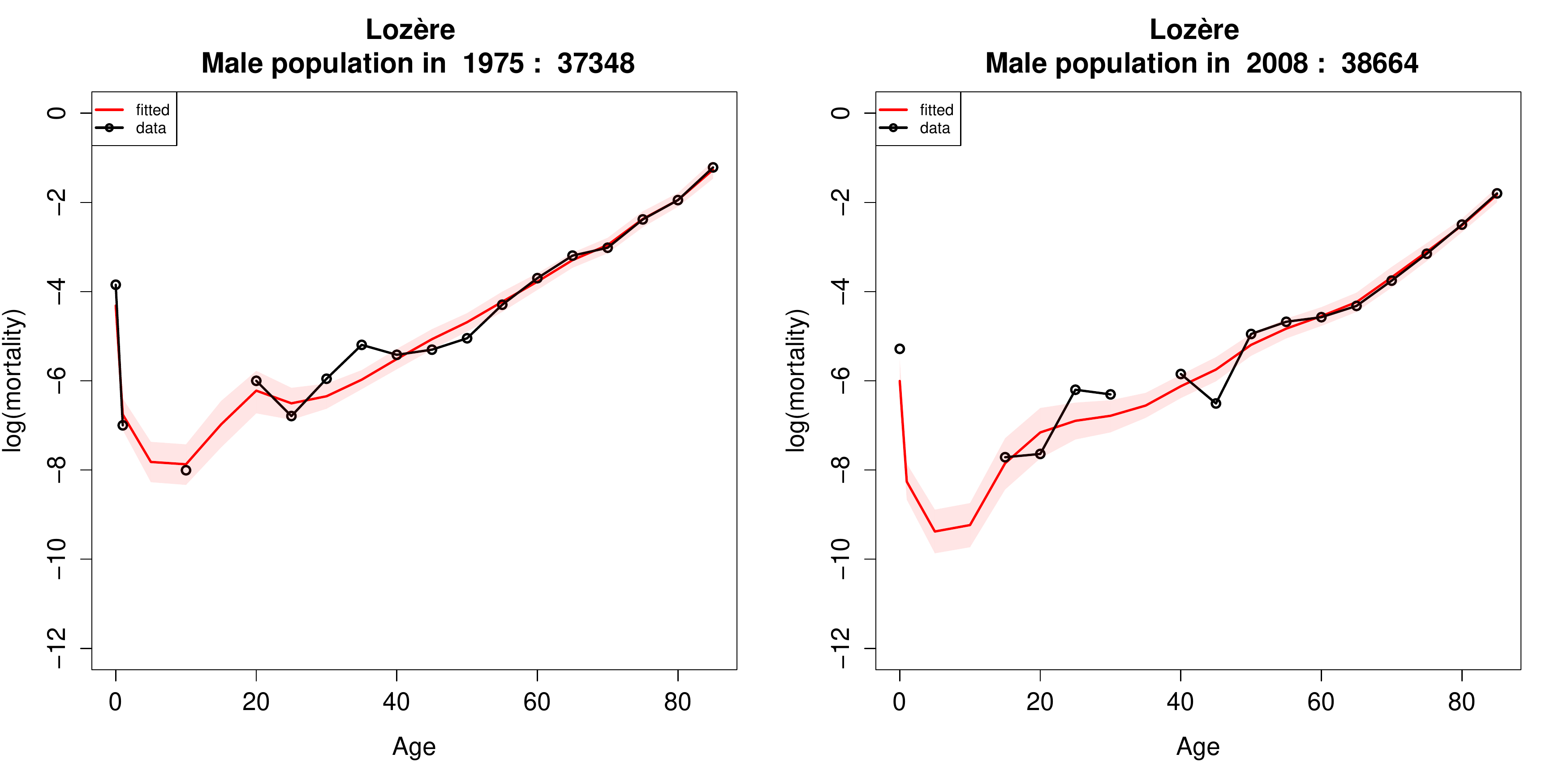}
\end{tabular}
\end{figure}

\begin{figure}[h!]
\centering
\caption{Observed and estimated log-mortality rates, Somme, Males, 1975 and 2008.}
\label{TEG}
\begin{tabular}{cc}
\includegraphics[width=0.9\textwidth]{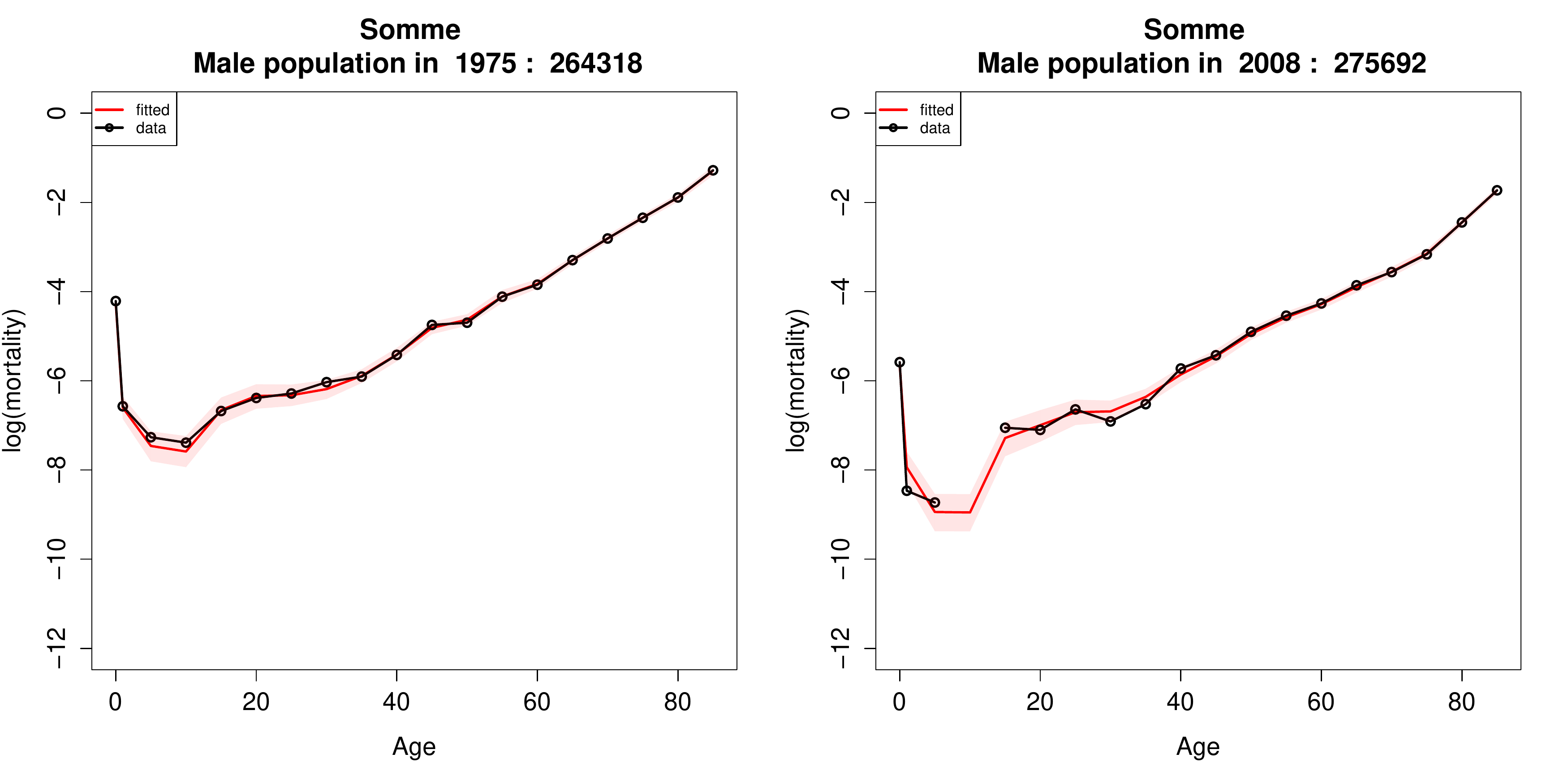}
\end{tabular}
\end{figure}

We used national France mortality curves from 1962--2008 to form the set of principal components used in estimation. For illustration, Figures \ref{MEL} and \ref{TEG} show the observed and estimated log-mortality rates for males in the departments Loz\`{e}re and Somme in 1975 and 2008. Loz\`{e}re has a male population of around 38,000, while Somme has a male population of around 275,000. For both \emph{d\'{e}partements}, there is a decrease in mortality rates from 1975 to 2008. As a consequence there are more zero death counts observed in 2008 compared with 1975, corresponding to more uncertainty around estimates in the more recent year. Additionally, there is less uncertainty around the Somme estimates, because the population size is around seven times the population in Loz\`{e}re.  

Once age-specific mortality rates have been estimated, other mortality measures and associated uncertainty can be calculated. Figure \ref{map} below shows  life expectancy at birth estimates for males in 2008. Life expectancy is estimated to be highest in areas around Paris and for the Midi-Pyrenees area, and lowest in the northern part of the country, a well-documented pattern (Barbieri, 2013). 

Uncertainty in life expectancy estimates is also easily obtained. Life expectancy is calculated for each of the posterior samples of age-specific mortality rates. A 95 per cent uncertainty interval is then obtained by calculating the $2.5^{th}$ and $97.5^{th}$ percentiles. For example, the estimate for life expectancy at birth for Paris in 2008 (both sexes) is 84.7 years (95\% UI: [84.5, 84.8]). For a smaller \emph{d\'{e}partements} such as Loz\`{e}re, the estimate is 81.9 years [81.5, 82.3]. Complete results for all French \emph{d\'{e}partements} are available in an online supplement to this paper.\footnote{See \texttt{http://shiny.demog.berkeley.edu/monicah/French/}.}

\begin{figure}[h!]
\caption{Life expectancy estimates for males, 2008 ($e_0$, years)}
\label{map}
\centering
\includegraphics[width=0.9\textwidth]{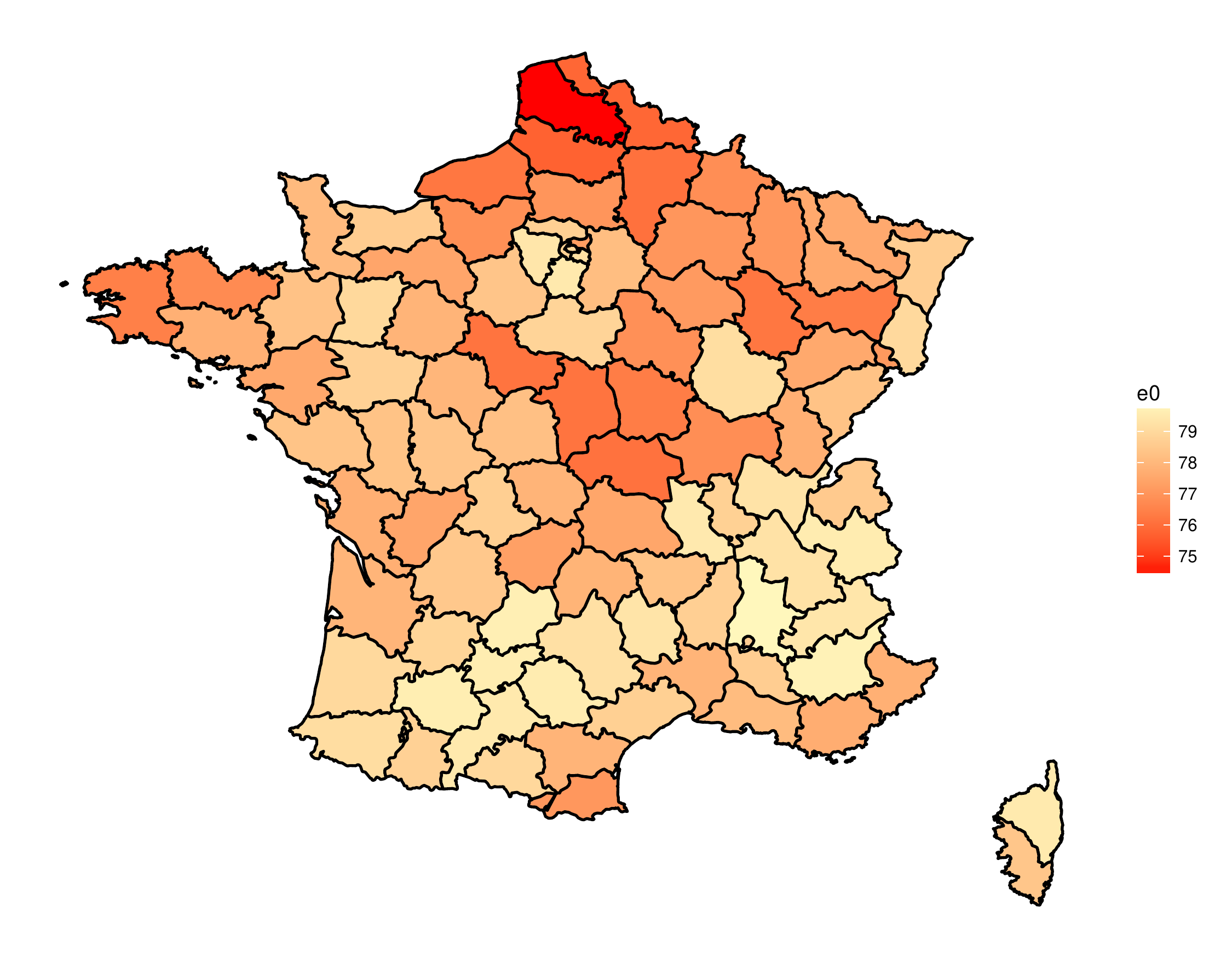}
\end{figure}

\begin{figure}[h!]
\caption{Variance of $\beta_1$ and $\beta_2$ over time}
\label{sigmas}
\centering
\includegraphics[width=0.9\textwidth]{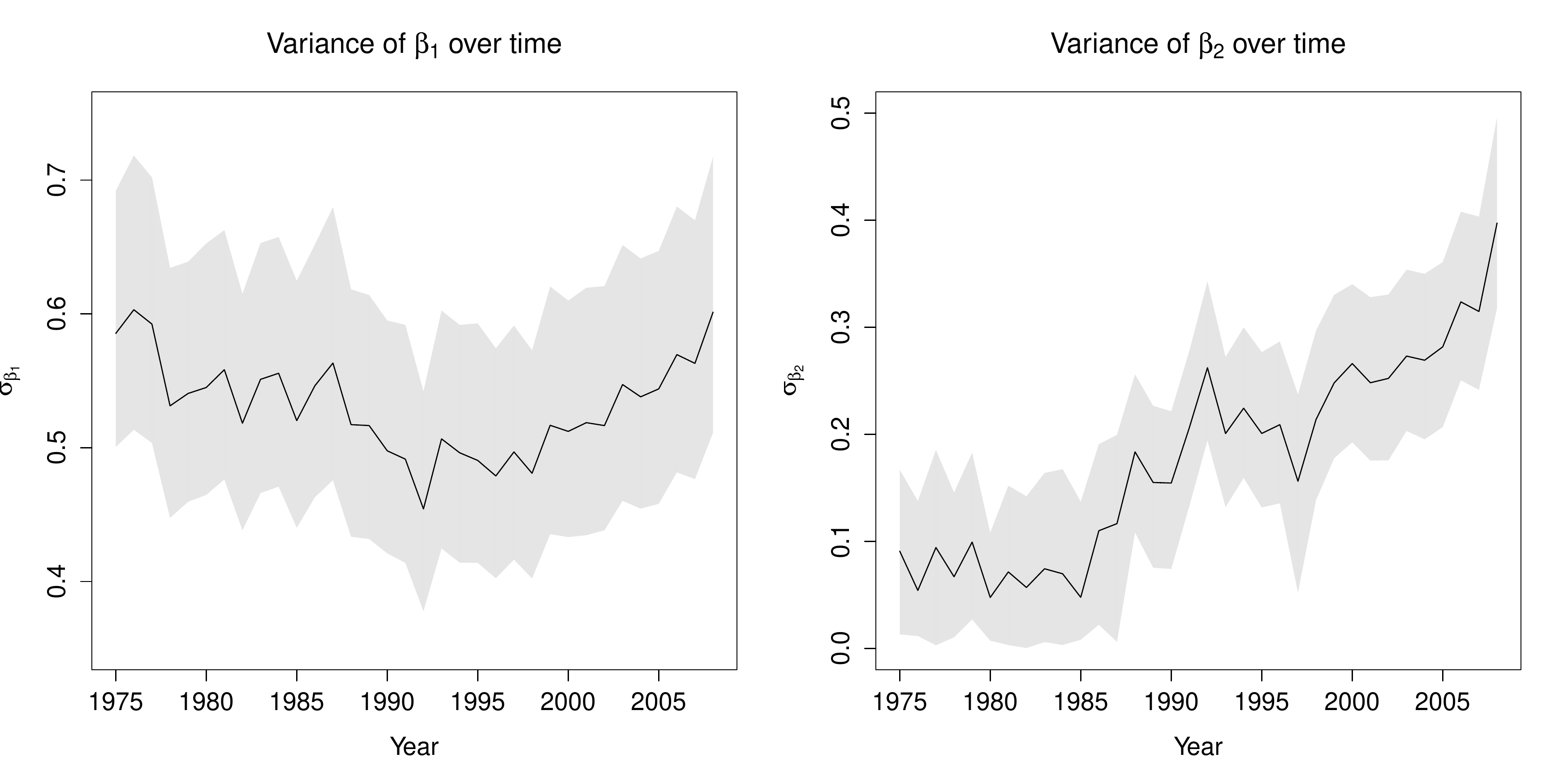}
\end{figure}

\begin{figure}[h!]
\caption{Low (left graph) and high (right graph) values of $\beta_2$}
\label{beta2comp}
\centering
\includegraphics[width=0.9\textwidth]{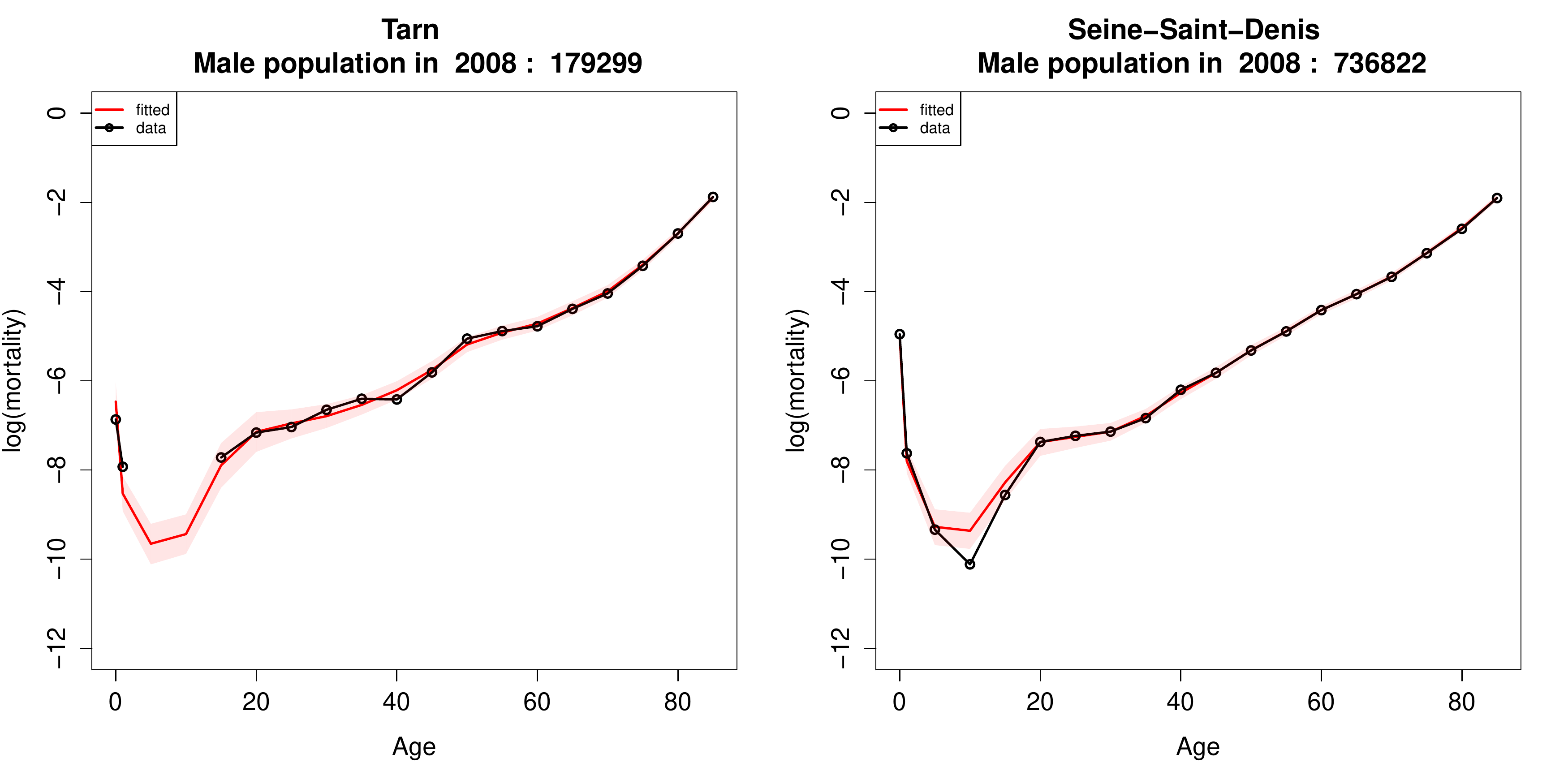}
\end{figure}

Another aspect of the results that could be of interest are the estimated variance parameters. It is assumed that the $\beta_{t,p}$ parameters are normally distributed with mean $\mu_{p,t}$ and variance $\sigma^2_{\beta_{p,t}}$. The variance terms may tell us something about how the spread of mortality outcomes is changing over time. Figure \ref{sigmas} shows the median and 95 per cent credible intervals of the variance parameters associated with  $\beta_1$ and $\beta_2$ over the period of estimation. While there is no discernible trend in $\sigma^2_{\beta_{1,t}}$, the variance parameter for the $\beta_2$ term, $\sigma^2_{\beta_{2,t}}$, appears to be increasing over time. The  $\beta_2$ related to the principal component which alters the relationship of the magnitude of infant and child mortality to mortality at older ages (see Figure \ref{pcs}). An increase in the variance parameter suggests that \emph{d\'{e}partements} are becoming more different over time with respect to child-versus-older mortality. Figure \ref{beta2comp} illustrates two \emph{d\'{e}partements} which have relatively low and high values of $\beta_2$. For Tarn, $\beta_2$ is low, which results in infant mortality being relatively low compared to adult mortality. For Seine-Saint-Denis, the opposite is true. 

\section{Conclusion}
We presented a novel method to estimate mortality rates by age and sex at the subnational level. In our approach we build on characteristic age patterns in mortality curves, pooling information across geographic space and smoothing over time within the framework of a Bayesian hierarchical model. When tested against simulated data, the model outperformed estimates from a simple Loess smoother and Brass model, especially for areas with smaller population sizes. The uncertainty in our estimates, reflected in the confidence intervals, is well calibrated. An application to real data for France illustrates how various parameter estimates from the model help to assess trends in overall mortality levels and inequalities within a country. The estimates produced by the model have direct applications to the study of subregional health patterns and disparities and how these evolve over time.

The model outlined in the paper is proposed as a general framework for estimating mortality rates in subpopulations. The framework can easily be altered by the user to best suit the situation in which rates are being estimated. For example, the mean parameter values $\mu_{\beta_p, t}$ need not be defined on a state/county basis, but may be defined by a smaller geographic area or based on some other characteristics, such as age distribution or rurality of an area. Additionally, it is possible to alter the random effects to be spatially structured, assuming some correlation in random effects by distance or of adjacent areas. 

The focus of this project has been on estimation of past and present mortality trends, rather than future ones. However, forecasting of age- and sex-specific mortality rates in a particular area can be obtained directly from model outputs. The mean parameters $\mu_{\beta_p, t}$ can be projected forward given the assumed linear time trend, which forms a basis to infer other parameters for areas of interest.  Given that the relevant variance parameters are also estimated in the process, uncertainty around forecasts can also be inferred. 

One of the contributions of our work is methodological. Estimates of mortality measures for small areas, most notably life expectancy, have been proposed previously. For example, Ezzati (2008) used information about number of deaths, together   with covariates related to socio-economic status, in order to estimate mortality rates. Congdon (2014) developed a random effects model to estimate life expectancy for subnational areas. Our method builds on some elements presented in the literature, while incorporating demographic knowledge about regularities in the Lexis surface of mortality rates by age and over time. More specifically, we complemented a state-of-the-art Bayesian hierarchical modeling framework to pool information across space and time, with a classic demographic approach to borrow information across age groups. In particular, our use of principal components of schedules of log-mortality rates is informed by a long tradition of demographic  modeling of mortality and can be considered an extension of the Lee-Carter approach (Lee and Carter 1992).   

In this article, we developed a general approach to complement classic demographic modeling ideas within a solid statistical framework. The model that we proposed and tested is quite minimalistic and relies on fairly simple rules for pooling across space and smoothing over time. However, the geographic scale at which spatial pooling will be implemented may depend on specific circumstances of different countries. Likewise, the type of time smoothing may vary. A number of rules can be accommodated within the framework that we propose.  

Potential extensions to the project include producing estimates by race or by cause of death for countries with high quality data. In the longer term, some of the ideas presented in the paper could be leveraged to generate estimates of mortality rates in the context of countries with low quality data.

\clearpage

\section*{References}

\begin{description}
\item Alkema, L and New, JR (2014). `Global estimation of child mortality using a Bayesian B-spline bias-reduction method.' \emph{The Annals of Applied Statistics} 8(4): 2122--2149.
\item Alkema, L, Raftery, AE, Gerland, P, Clark, SJ, Pelletier, F, Buettner, T, Heilig, GK (2012). `Probabilistic Projections of the Total Fertility Rate for All Countries', \emph{Demography}, 48(3), 815--839. 
\item Barbieri, M (2013). `Mortality in France by \emph{d\'{e}partements}', \emph{Population}, Vol. 68, 2013(3): 375--417.
\item Bravo, JM, Malta (2010). `Estimating life expectancy in small population areas.' Work session on demographic projections, EUROSTAT-EC Collection: Methodologies and working papers, Theme: Population and Social Conditions: 113-126.
\item Bijak, J. (2007) `Bayesian Methods in International Migration Forecasting', in J Raymer and F Willekens (eds): \emph{International Migration in Europe: Data, Models and Estimates}, John Wiley \& Sons, Ltd, Chichester, UK.
\item Chetty, R, Stepner, M, Abraham, S, Lin, S, Scuderi, B, Turner, N, Bergeron, A, and Cutler, D (2016). `The Association Between Income and Life Expectancy in the United States, 2001--2014', \emph{Journal of the American Medical Association}, 315(16):1750--1766.
\item Congdon P. (2009) Life expectancies for small areas: a Bayesian random effects methodology. \emph{Int Stat Rev}, 77(2): 222--240.
\item Congdon P. (2014) Estimating life expectancies for US small areas: a regression framework. \emph{J Geogr Syst }, 16: 1-18.
\item Currie, J and Schwandt, H (2016). `'Mortality Inequality: The Good News from a County-Level Approach', \emph{Journal of Economic Perspectives}, 30(2): 29--52. 
\item Ezzati, M, Friedman, AB, Kulkarni, SC and Murray, CJL (2008). The reversal of fortunes: trends in county mortality and cross-county mortality disparities in the United states. \emph{Plos Med}, 5:e66.
\item Gelman, A. and Rubin, D. (1992). Inference from iterative simulation using multiple sequences. \emph{Statist. Sci.} 7:457--511.
\item Girosi, F and King, G (2008). \emph{Demographic Forecasting}. New Jersey: Princeton University Press. 
\item  Institut national de la statistique et des \'{e}tudes \'{e}conomiques (INSEE) (2015). French department mortality data; personal communication to Magali Barbieri by the {Division des statistiques r\'{e}gionales, locales et urbaines}, INSEE. 
\item Jarner, SF and Kryger, EM (2011) `Modelling Adult Mortality in Small Populations: The SAINT Model.' \emph{Astin Bulletin} 41(2): 377-418.
\item Kindig DA and Cheng ER (2013): `Even as mortality fell in most US counties, female mortality nonetheless rose in 42.8 Percent of counties from 1992 to 2006.' \emph{Health Aff (Millwood)}, 32(3):451--8.
\item King, G, and Soneji, S (2011). `The future of death in America', \emph{Demographic Research}, 29, Article 1. 
\item Kulkarni SC, Levin-Rector A, Ezzati M, Murray CJ (2011). `Falling behind: life expectancy in US counties from 2000 to 2007 in an international context.' \emph{Popul Heal Metrics} 9:16.
\item Lee, RD and Carter, LR (1992). Modeling and forecasting U.S. mortality. \emph{Journal of the American Statistical Association} 87(419): 659--675.
\item Plummer, M (2003). `JAGS: a program for analysis of Bayesian graphical models using Gibbs Sampling.' In Proceedings of the 3rd International Workshop on Distributed Statistical Computing (DSC 2003), March 20-22, Vienna, Austria. ISSN 1609-395X. Available at: \verb|http://mcmc-jags.sourceforge.net/|.
\item Macintyre S, Ellaway, A and Cummins, S (2002): `Place effects on health: how can we conceptualise, operationalise and measure them?', \emph{Social Science \& Medicine}, 55: 125--139.
\item Raftery, AE, Li, N, Sevcikova, H, Gerland, P, and Heilig, GK (2012). `Bayesian probabilistic population projections for all countries.' \emph{Proceedings of the National Academy of Sciences of the USA}, 109, 13915--13921.
\item Schmertmann, C, Zagheni, E, Goldstein, J and Myrskyla, M (2014). Bayesian Forecasting of Cohort Fertility, \emph{Journal of the American Statistical Association}, 109(506). 
\item Sharrow, DJ, Clark, SJ, Collinson, MA, Kahn, K, and Tillman, SM (2013). `The age pattern of increases in mortality affected by HIV: Bayesian fit of the Heligman-Pollard Model to data from the Agincourt HDSS field site in rural northeast South Africa', \emph{Demographic Research}, 29, Article 39. 
\item Srebotnjak T, Mokdad, AH and Murray, CJL (2010). \emph{Population Health Metrics}, 8:26.
\item United States Census Bureau (USCB) (2015). `County Totals. Vintage 2014.' Available at: \verb|https://www.census.gov/popest/data/counties/totals/2014/index.html|. 
\end{description}

%
%
%

\end{document}